\documentclass[10pt, conference]{IEEEtran}

\usepackage{adjustbox}
\usepackage{subcaption}
\usepackage{caption}
\usepackage{graphicx} 
\usepackage{stfloats}
\usepackage{booktabs}
\usepackage{multirow}
\usepackage{optidef}
\usepackage{subcaption}

\IEEEoverridecommandlockouts
\usepackage[percent]{overpic}
\usepackage{graphicx}
\usepackage{amsmath,amssymb,amsfonts}
\usepackage{float}
\usepackage{multicol}
\usepackage{lipsum} 
\usepackage{textcomp}
\usepackage{xcolor}
\usepackage{breakurl}
\usepackage{xurl}
\usepackage{longtable} 
\usepackage{array}     
\usepackage{verbatim}
\usepackage[ruled, vlined, linesnumbered, boxed]{algorithm2e}
\usepackage{algpseudocode}

\title{Round Trip Time: A Benign Signal or an Indirect Window into Datacenter Workloads?\thanks{This material is based upon work supported by the National Science Foundation (NSF) under Award Number CNS-2232889.}}

\author{
Sourya Saha, Md Nurul Absur, Saptarshi Debroy\\
City University of New York\\ 
Emails: \textit{ssaha2@gradcenter.cuny.edu, mabsur@gradcenter.cuny.edu, saptarshi.debroy@hunter.cuny.edu}
}

\begin{document}

\maketitle
\thispagestyle{empty}
\pagestyle{empty}
\maketitle

\begin{abstract}
Multi-tenant datacenter networks increasingly rely on shared leaf-spine fabrics, where traffic from multiple tenants traverses common network resources. While logical isolation mechanisms prevent direct access between tenants, shared congestion dynamics may still expose indirect information about co-located workloads through observable latency variations. In this paper, we investigate a network side-channel vulnerability arising from shared congestion behavior in multi-tenant datacenter fabrics using RTT observations collected along overlapping network paths. We develop a framework to explore how workload-induced latency variations contain sufficiently distinguishable signatures to enable workload inference under realistic deployment conditions. Our evaluations show that indirect RTT observations can reveal meaningful workload information, achieving up to 97.3\% run-level accuracy under cross-path evaluation when workload-induced congestion is sufficiently observable. The findings suggest that logical network isolation alone may be insufficient to prevent information leakage through shared congestion dynamics in modern datacenter infrastructures.

\end{abstract}

\begin{IEEEkeywords}
Round trip time, workload inference, network side channel, datacenter network, leaf-spine topology, congestion.
\end{IEEEkeywords}

\vspace{-0.1in}
\section{Introduction}
\label{sec:intro}

Modern datacenter networks have converged on multi-tenant shared infrastructure driven by cloud economics, where multiple tenants (sometimes adversarial) share the same physical fabric. The dominant leaf-spine topology connects compute nodes through leaf switches and interconnects leaves via spine switches \cite{al-fares2008, greenberg09, singh15}. While this design provides high network capacity and predictable latency, it also means that the traffic from different tenants traverses shared spine switches, creating observable congestion points. As workloads increasingly include sensitive applications such as financial transactions and medical analytics, the risk of information leakage increases \cite{9754271}. Unlike compute isolation, which benefits from mature mechanisms such as virtual machines, containers, and trusted execution environments, network isolation remains largely logical, (e.g., VLANs), without sufficient physical separation. Although such logical separation prevents direct access, it does not mitigate indirect leakage through shared resource contention - a precursor to network side-channel vulnerabilities. At scale, this exposure is significant. Major providers host millions of tenants on shared fabrics, and similar designs are increasingly adopted in private datacenters \cite{popa2012,multi-cloud,adon,icnp}. 

\begin{figure}[tb]
  \centering
  \includegraphics[width=\columnwidth]{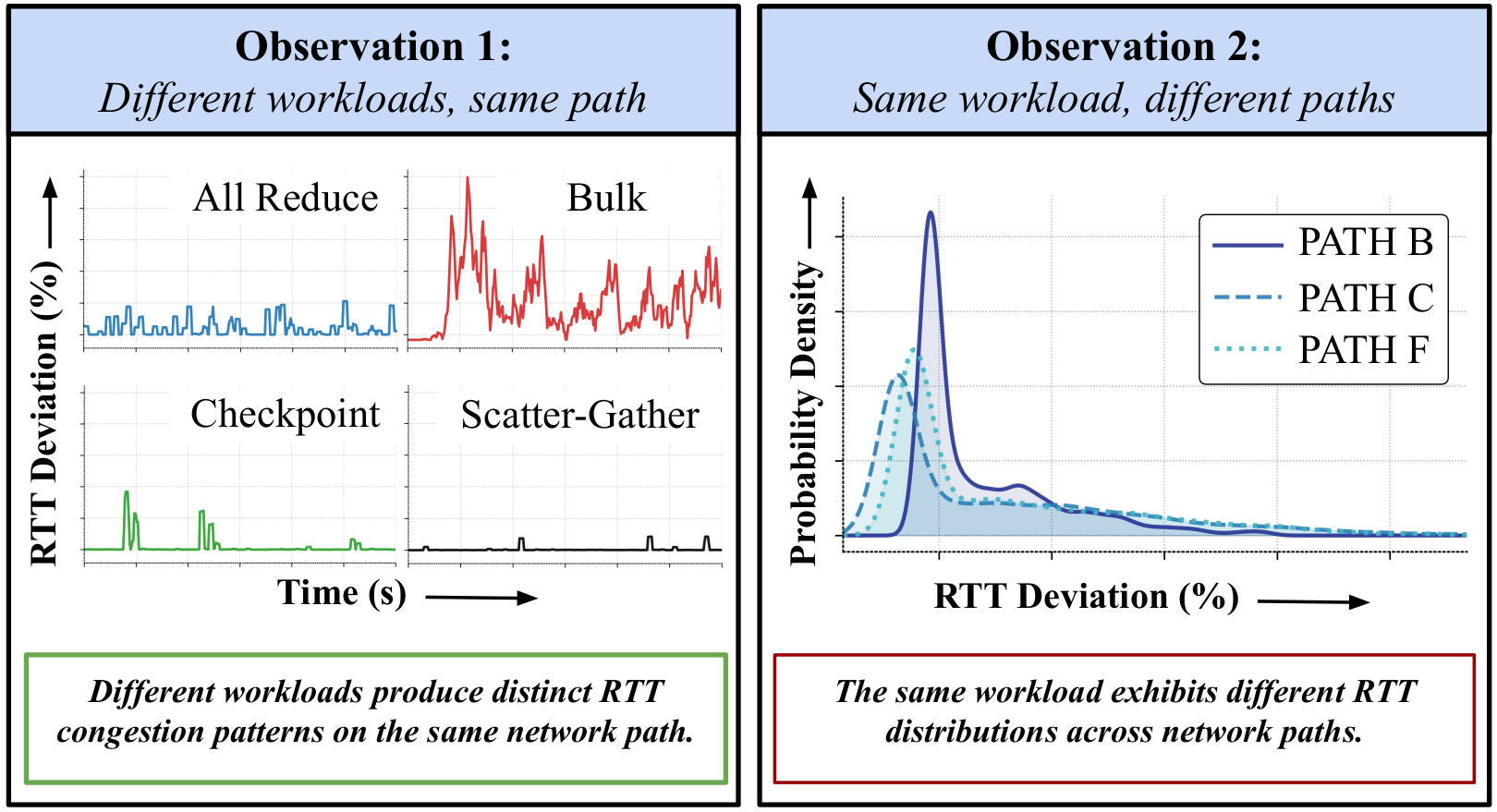}
  \caption{RTT side-channel signal characteristics. (Left) RTT deviation traces (60-second window, 5-sample rolling average) for four workloads on the same network path, showing distinct congestion patterns. (Right) Kernel density estimation of RTT deviations for \emph{bulk transfer} workload across three network paths, illustrating cross-path distribution shift.}
  \label{fig:problem}
  \vspace{-0.2in}
\end{figure}

Building on such shared network infrastructure, a natural question arises: \emph{can the activity of one tenant be inferred by another through indirect network effects?} When multiple workloads share network resources, the traffic generated by one can influence shared components such as queues and buffers, leading to measurable variations in end-to-end latency along overlapping paths. In principle, such variations may carry information about underlying workload activity. However, extracting meaningful information from these signals is fundamentally challenging. The relationship between workload behavior and observed latency is indirect and shaped by multiple factors, including background traffic, buffering dynamics, and network control mechanisms, which can obscure or distort the signal. In addition, the same workload may manifest differently under varying network conditions, making it difficult to distinguish consistent patterns, as illustrated in Fig. \ref{fig:problem}. 
Finally, the magnitude of latency variations may be small relative to natural fluctuations in the network, further complicating reliable interpretation. Together, these challenges make it unclear whether such indirect observations can support robust workload inference in practical multi-tenant datacenter environments~\cite{vectrust}.

Prior work has shown that shared infrastructure can leak information through contention effects, including timing variations in shared compute resources and limited forms of network-level inference in multi-tenant environments \cite{zhang2025nvbleed,wu_netcovert,bates2012_cloudnetsidechannel}. Existing explorations generally rely on either direct visibility into shared hardware state or access to packet-level traffic features for inference \cite{shapira2021_flowpic,measurement,pca}. However, far less is understood about whether workload activity can be inferred purely through indirect congestion effects in shared datacenter fabrics, where observations are weak, aggregated, and shaped by path-dependent network dynamics. This raises a broader question of whether shared network infrastructure itself can expose distinguishable workload signatures even without access to victim traffic or local hardware measurements.

In this paper, we investigate a network side-channel vulnerability that allows inferring workload activity in multi-tenant leaf-spine datacenter fabrics using latency observations along shared network paths. The key idea is that shared network resources introduce indirect signals in end-to-end delay that reflect underlying workload behavior, even in the absence of direct traffic visibility. To investigate this vulnerability, we develop three key components. First, we design a set of path-invariant temporal features that capture workload-induced patterns in latency while remaining robust to differences in baseline network conditions. Second, we develop a domain adaptation strategy that enables models trained under one set of network conditions to generalize across others by aligning feature distributions and filtering path-dependent noise. Third, we build a streaming inference framework that operates continuously on latency observations to produce workload predictions in real time. Together, these components enable meaningful workload inference from indirect and noisy signals. Our findings suggest that logical isolation alone may be insufficient to prevent information leakage through shared network effects, and that even limited external observations can reveal information about co-located workloads.

We evaluate the feasibility of such side channel vulnerabilities on a controlled multi-site datacenter testbed deployed on the FABRIC infrastructure \cite{fabric-2019}, using a geographically distributed leaf-spine topology with controlled link capacities. Our evaluation spans a diverse set of representative datacenter workloads and considers scenarios where observations are collected under different network conditions. We design the study to capture variability in path characteristics and background traffic, and to assess robustness across deployment settings, achieving 73.7\% run-level accuracy under cross-path evaluation, rising to 97.3\% when workloads with weak congestion signatures are grouped into a single low-signal class. The evaluation further examines the effect of observation duration and workload characteristics on inference performance, along with the inherent limitations imposed by weak or noisy signals. Together, this setup enables a comprehensive assessment of the feasibility and boundaries of workload inference from indirect network observations.

The rest of the paper is organized as follows. Section \ref{sec:related-works} reviews related work. Section \ref{sec:threat-model} describes the Vulnerability model and presents the formulation of the problem. Section \ref{sec:solution-strategy} discusses our inference process design. Section \ref{sec:eval} details the implementation, evaluation, and results. Finally, Section \ref{sec:conclusion} concludes the paper and discusses the future work.


\section{Related Work}
\label{sec:related-works}

Existing work relevant to this vulnerability can be grouped into three broad categories. The first major range includes studies of side-channel leakage in shared hardware and cloud infrastructure. They make the case for the fact that logical isolation does not eliminate leakage when adversaries and victims contend for common physical resources  \cite{10.1145/1653662.1653687,8241986,sanjaya2025applicationspecificpowersidechannelattacks}.  The second set of studies focuses on traffic analysis and website fingerprinting, flow-level patterns such as packet timing, and burst structure. These studies describe how these patterns are used to infer sensitive information even when payload contents are hidden \cite{pekar2026tutorialflowbasednetworktraffic,li2025wifingerfingerprintingnoisyiot}.  The third set of studies addresses the shift in distribution through domain adaptation, which becomes essential when profiling and deployment conditions differ in scenarios
\cite{sun2016coral,wu2024graphlearningdistributionshifts}. Our contribution lies at the intersection of these areas, but is significantly different from each in a primary level. We do not assume access to rich local contention signals, unlike traditional hardware side-channel literature. Our work also differs from the current work in that it does not assume visibility into the victim’s packets or flow sequence. Lastly, our source of the distributional shift is induced by path-dependent network dynamics in a shared leaf-spine topology.

\subsubsection{Side Channels in Shared Hardware and Cloud Infrastructure}

Many significant works have shown that shared infrastructure can leak information through contention effects. Traditional works mainly capitalized CPU caches and related micro-architecture state \cite{281354,adiletta2025spillbeansexploitingcpu}. Relatively recent work such as \cite{11424401,10.1145/3793532} extended the leakage surfaces to GPU, DNN accelerators, and multi-GPU interconnects. These works demonstrate that timing variation and shared-resource behavior can support application fingerprinting and inference. In cloud infrastructure, co-residency studies have established that multi-tenancy itself can be a viable path to leakage. The cloud infrastructure creates stronger isolation mechanisms at both the compute and network layers  \cite{11158758}. However, these works typically rely on the assumption of either host-level co-location or access to high-resolution local measurements directly tied to the victim’s execution environment. 

\vspace{-0.452mm}
  
\subsubsection{Traffic Analysis, Encrypted Traffic Classification, and Network Side Channels}

Several studies have shown that application behavior can often be inferred from encrypted traffic using packet timing, burst patterns, and flow-level structure \cite{feng2025unmasking,9896143, MEHAVILLA2026115893}. More recent contributions have correspondingly treated packet timing and size as side-channel carriers that require mitigation \cite{pulkit2023surveysidechannelattackscontext, 10.1145/3645109}. The works aforementioned demonstrate that network observations can reveal sensitive information even when the content is inaccessible. Existing state-of-the-art methods typically assume direct access to the victim’s packet stream and use packet-level features such as size, direction, and inter-arrival structure \cite{meghdouri2023machine, 11455356}. However, such access to information certainly hampers the fairness of detection.

Based on the current literature, side channels in shared hardware and cloud infrastructure mostly rely on prior assumptions. This makes the signal more aggregated, noisier, and more sensitive to deployment conditions. Also, existing methods usually assume direct access to the victim packet stream, which is unrealistic in real-world deployments. \emph{In contrast, our work only observes RTT perturbations of its own probes to the victim, appearing indirectly through shared queuing dynamics without any prior assumptions. Our evaluation further demonstrates robustness across changes in path characteristics, background traffic, and buffering behavior. To the best of our knowledge, our work is the first one to comprehensively address these research gaps and with both qualitative \& quantitative evaluations. Overall, our contribution is not simply to apply classification to RTT traces, but to frame workload inference from indirect network-side effects as a transfer-robust learning problem using weak proxy observations.}


\section{Vulnerability Model and Problem Formulation}
\label{sec:threat-model}

\subsection{System Model}

\begin{figure}[tb]
  \centering
  \includegraphics[width=\columnwidth]{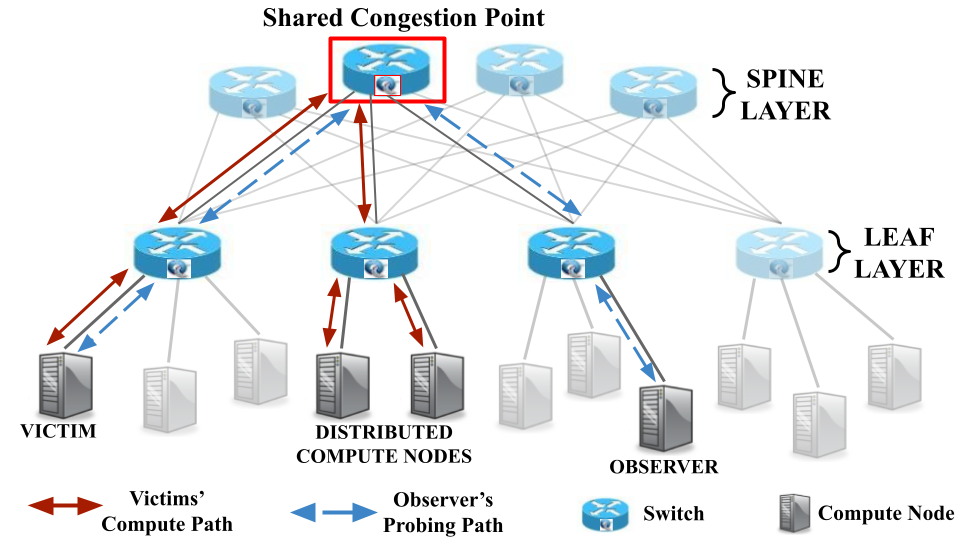}
  \caption{\textbf{Our system model: } Leaf-spine topology with shared spine switch (highlighted at the top). The victim's workload traffic ({\color{red}red}) and observer's RTT probes ({\color{blue}blue dashed}), traverse the same spine, creating a congestion side-channel. The observer observes only their own probe RTTs; it has no access to victim traffic."}
  \label{fig:topology}
    \vspace{-0.5cm}
\end{figure}

We consider a multi-tenant datacenter network organized in a \emph{leaf-spine (Clos) topology}, which is widely used in modern large-scale deployments. In this architecture, leaf switches connect compute nodes, while spine switches interconnect all leaves. Traffic between nodes attached to different leaves traverses a two-hop path via a single spine switch:
\[
\text{source} \rightarrow l_i \rightarrow s_j \rightarrow l_k \rightarrow \text{destination}
\]
where $l_i, l_k \in \mathcal{L}$ are leaf switches and $s_j \in \mathcal{S}$ is the selected spine. Figure~\ref{fig:topology} illustrates this architecture and highlights the shared forwarding path through the spine layer.
A key property of this architecture is that spine switches act as shared resources. Traffic from multiple tenants that is mapped to the same spine contends for common output buffers, creating shared congestion points. As a result, variations in one tenant’s traffic can influence the latency experienced by other flows traversing the same spine, even without direct interaction between tenants.

\vspace{-0.1in}
\subsection{Adversarial Setup}

We define the observer as a tenant controlling a set of compute nodes $\mathcal{A} = {a_1, a_2, \ldots, a_m}$ connected to the datacenter fabric. The observer has standard network access and can send probe traffic to any reachable node within the fabric, as well as measure end-to-end latency of these probes. The observer may also execute arbitrary workloads on nodes they control for offline profiling, but has no access to the victim’s nodes, traffic, or network infrastructure.

We assume that the observer’s probe traffic shares at least one network component with the victim’s traffic, enabling indirect observation through shared resource contention. The observer cannot intercept or inspect victim packets, and relies solely on latency variations induced by shared network effects. We further assume that the victim runs a dominant workload over a sufficient observation period, allowing stable patterns to emerge in the observed signal. In addition, the observer is assumed to have prior knowledge of candidate workload classes obtained through separate profiling under similar conditions. Accordingly, we consider a closed-set inference setting, where the framework distinguishes among known workload categories rather than previously unseen workloads.

Under this model, probing is performed from a selected node $a \in \mathcal{A}$, producing a time series of round-trip times $\mathbf{r}_a = [r_a(t_1), r_a(t_2), \ldots, r_a(t_N)]$. Each measurement reflects a combination of fixed path latency, congestion-induced delay, and measurement noise. While constant components remain stable, the time-varying portion of the signal is influenced by workload-driven network activity, which introduces observable patterns in latency over time. The observer’s objective is to infer the underlying workload class from these patterns, despite the signal being indirect and noisy. We formalize this inference problem and its associated challenges in the following subsection.

\vspace{-0.1in}
\subsection{Problem Formulation}

We formalize workload inference as a classification problem under indirect observation. The observer observes a time series of RTT measurements $\mathbf{r} = [r(t_1), r(t_2), \ldots, r(t_N)]$ collected over an observation period, and aims to infer the workload label $\hat{y} \in \mathcal{W} = \{w_1, \ldots, w_C\}$ from a set of candidate classes. The observer has access to RTT traces collected during prior profiling, where each trace is associated with a known workload, forming a dataset $\mathcal{D}^{(\text{prof})} = \{(\mathbf{r}_{c,k}^{(\text{prof})}, w_c)\}$.

A key challenge arises from differences between the conditions under which profiling data is collected and those under which inference is performed. As a result, the distribution of observed RTT traces may differ across these settings, such that $P^{(\text{prof})}(\mathbf{r} \mid y) \neq P^{(\text{vic})}(\mathbf{r} \mid y)$. The observer may also collect additional RTT traces along the deployment path, but without knowledge of the corresponding workload.

The objective is to learn a mapping $h: \mathcal{R} \to \mathcal{W}$ that generalizes across these conditions:
\[
h^* = \arg\max_{h \in \mathcal{H}} \; \mathbb{E}_{(\mathbf{r}, y) \sim P^{(\text{vic})}} \left[ \mathbb{1}[h(\mathbf{r}) = y] \right]
\]
where $\mathcal{R}$ denotes the space of RTT time series, $\mathcal{H}$ is the hypothesis class of candidate classifiers, $P^{(\text{vic})}$ represents the distribution of observations under deployment conditions, and $\mathbb{1}[\cdot]$ is the indicator function that evaluates to 1 when the prediction is correct and 0 otherwise.
Solving this problem requires extracting features from RTT signals that capture workload-dependent patterns while remaining robust to path variations, adapting representations across environments, and learning a classifier that operates on indirect and noisy observations.

\vspace{-2.8mm}
\subsection{Vulnerability Model}


Our vulnerability model operates in two phases: \emph{profiling} and \emph{inference}.

In the \emph{profiling} phase, the observer executes each target workload $w_c \in \mathcal{W}$ on nodes they control while simultaneously collecting RTT measurements along a probe path that traverses the shared network fabric. As seen in Fig. \ref{fig:topology}, the observer might choose to use any other nodes under its control (depicted as translucent compute nodes attached to different leaves) for this purpose. For each workload, this produces a set of time series $\mathbf{r}^{(\text{prof})}$ that capture latency variations under known workload conditions. Since absolute RTT values depend on path-specific factors such as propagation delay, the observed signal is normalized relative to its baseline:
\[
\tilde{r}(t_i) = \frac{r(t_i) - \hat{r}}{\hat{r}}, \quad \hat{r} = \mathrm{median}(\mathbf{r})
\]
This normalization removes constant path-dependent components and expresses the signal as a relative deviation, isolating time-varying effects associated with shared network contention.

In the \emph{inference} phase, the observer collects RTT measurements along a path that overlaps with the victim’s traffic, producing a time series $\mathbf{r}^{(\text{vic})}$. The same normalization is applied:
\[
\tilde{r}^{(\text{vic})}(t_i) = \frac{r^{(\text{vic})}(t_i) - \hat{r}^{(\text{vic})}}{\hat{r}^{(\text{vic})}}
\]
The resulting normalized time series is then used as input to an inference procedure that maps observed latency patterns to workload classes.


\vspace{-0.15in}
\section{Inference Strategy}
\label{sec:solution-strategy}

The workload inference problem defined in Section \ref{sec:threat-model} presents several fundamental challenges, including extracting meaningful patterns from indirect and noisy RTT signals, ensuring robustness across different network paths, and capturing workload behavior that spans multiple temporal and structural scales.
To address these challenges, our work adopts a structured approach that transforms raw RTT observations into path-independent representations, aligns distributions across deployment conditions, and enables reliable inference over time. The design is guided by key observations about the characteristics of RTT-based side-channel signals, which we present next.

\begin{figure}[tb]
  \centering
  \includegraphics[width=0.85\columnwidth]{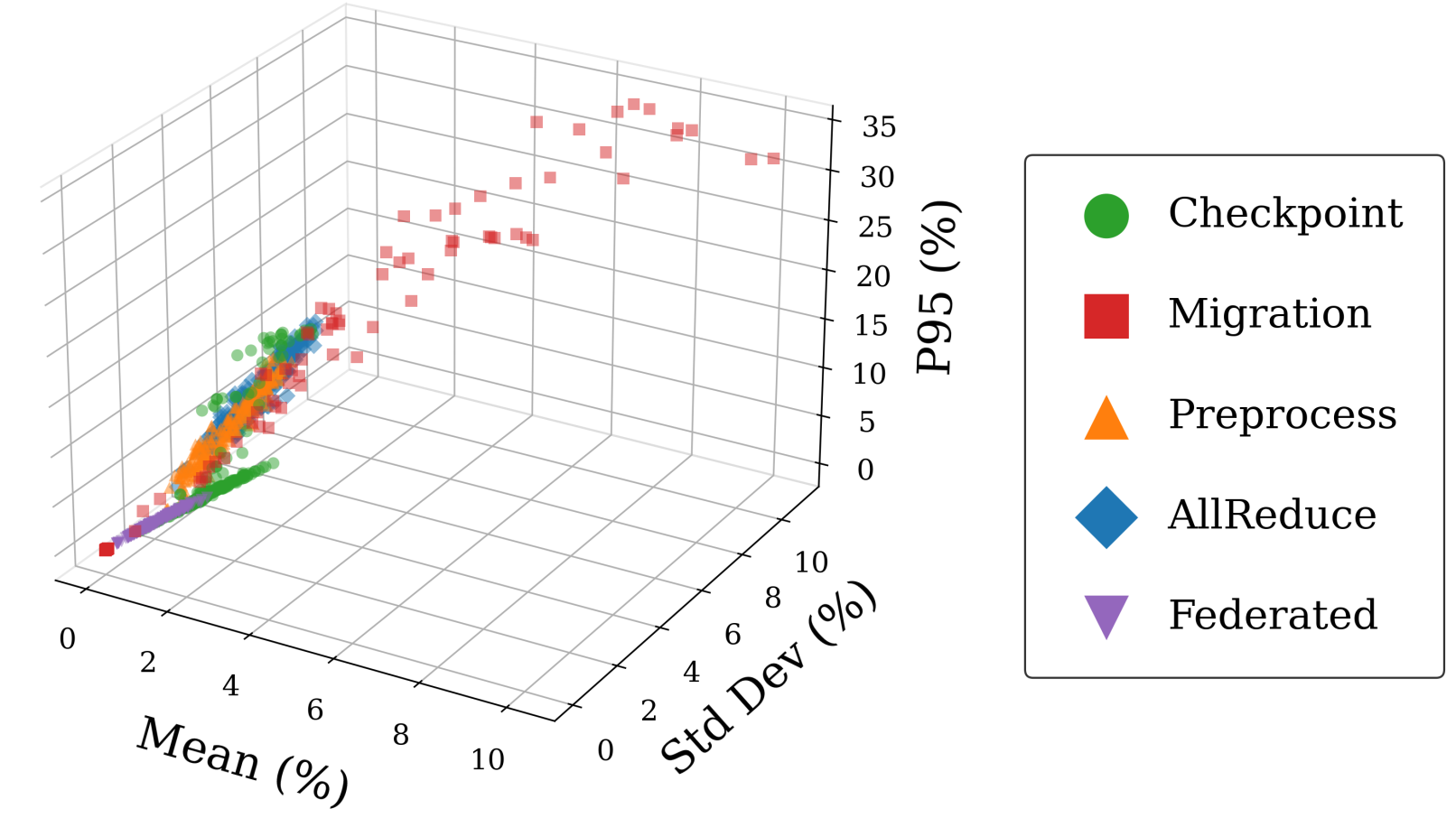}
  \caption{Per-window mean, standard deviation, and 95th percentile of normalized RTT deviation for  
  five workloads (60-second windows, Path B, 10 runs each). The extensive overlap across workloads   
  indicates that these features alone are inadequate for reliable classification.}
  \label{fig:motivation}
  \vspace{-0.2in}
\end{figure}

\subsection{Motivation: Why Raw RTT Features Fail}
\label{sec:rtt-fail}
A natural starting point is to apply standard traffic-classification features, such as mean, variance, and percentiles, directly to the RTT time series. However, these features provide limited discrimination for the RTT side-channel signal. As shown in Figure \ref{fig:motivation}, windows from five distinct workloads occupy heavily overlapping regions when projected onto the mean, standard deviation, and 95th-percentile feature space, making reliable classification difficult. Table \ref{tab:design-observations} summarizes the key signal characteristics that motivate our subsequent design choices.

\begin{table}[htbp]
\centering
\caption{Key Signal characteristics motivating the inference design.}
\label{tab:design-observations}
\scriptsize
\setlength{\tabcolsep}{3pt}
\renewcommand{\arraystretch}{1.12}

\begin{tabular}{
@{}
>{\raggedright\arraybackslash}p{0.19\columnwidth}
>{\raggedright\arraybackslash}p{0.45\columnwidth}
>{\raggedright\arraybackslash}p{0.29\columnwidth}
@{}
}
\toprule
\textbf{Observation} & \textbf{Challenge} & \textbf{Design implication} \\
\midrule

\textbf{O1: Path-dependent baseline}
&
The measured RTT $r(t)$ is dominated by the path-specific propagation delay $d_{\mathrm{prop}}$, while the workload-induced queuing variation $\Delta d_{\mathrm{queue}}(t)$ appears only as a small perturbation. Residual feature correlations may also vary across paths after normalization.
&
Construct path-independent representations and apply cross-path adaptation.
\\

\addlinespace[2pt]

\textbf{O2: Temporal scale}
&
Workload signatures range from sub-second bursts to long-duration periodic behavior. Fixed short windows may fragment or miss informative structures that span longer timescales.
&
Preserve workload-dependent temporal structure across multiple timescales.
\\

\addlinespace[2pt]

\textbf{O3: Feature-domain diversity}
&
Workloads differ in amplitude, burstiness, periodicity, persistence, and low-intensity temporal behavior. No single feature domain captures all of these signatures.
&
Combine complementary statistical, temporal, spectral, and long-range dependence features.
\\

\bottomrule
\end{tabular}
\vspace{-0.2in}
\end{table}

\subsection{Path-Independent Feature Design}

Guided by the observations in Section \ref{sec:rtt-fail}, we design a feature extraction pipeline that satisfies two key requirements: \emph{path independence}, so that representations transfer across network paths, and \emph{multi-scale discrimination}, so that workload patterns are captured across a wide range of temporal and traffic regimes.

\subsubsection{Normalization}

All features are computed on a normalized RTT signal:
\[
\tilde{r}(t_i) = \frac{r(t_i) - \hat{r}}{\hat{r}}, \quad \hat{r} = \text{median}(\mathbf{r}_{\text{window}})
\]
This transformation removes path-dependent baseline effects by expressing RTT as a fractional deviation from its local baseline. As a result, the representation captures relative congestion dynamics rather than absolute latency, enabling comparability across paths with different propagation delays.

\subsubsection{Feature Representation}

From the normalized signal, we extract a fixed-dimensional feature vector using a mapping $\phi(\cdot)$ that captures \emph{amplitude structure and congestion variability}, \emph{temporal burst and event patterns}, \emph{multi-scale variability across time}, \emph{periodicity and frequency structure}, \emph{long-range dependence and persistence}, and \emph{temporal complexity in low-intensity signals}. Together, these complementary feature groups form a multi-domain representation that captures workload-specific behavior beyond any individual feature domain.

\vspace{-2mm}

\subsection{Cross-Path Domain Adaptation}

Path-independent features reduce but do not eliminate the cross-path domain gap, as feature correlations and noise characteristics may still vary with background traffic and buffering conditions. We therefore employ a two-stage adaptation pipeline as follows.

\subsubsection{Dimensionality Reduction}

We apply Principal Component Analysis (PCA) to project the standardized feature space onto a compact subspace that captures the majority of the variance while suppressing noise and redundancy. PCA is preferred over explicit feature selection because discriminative information is distributed across multiple feature domains; retaining this information produces a representation that is less sensitive to path-specific variation and overfitting.

\subsubsection{Covariance Alignment (CORAL)}

After dimensionality reduction, we align the second-order statistics of the profiling features to match those observed along the deployment path using CORAL \cite{sun2016coral}:
\[
\hat{\mathbf{x}} = (\mathbf{x}^{(\text{prof})} - \boldsymbol{\mu}^{(\text{prof})}) \cdot (\boldsymbol{\Sigma}^{(\text{prof})})^{-1/2} \cdot (\boldsymbol{\Sigma}^{(\text{vic})})^{1/2} + \boldsymbol{\mu}^{(\text{vic})}
\]
This transformation whitens the profiling features and re-colors them using the covariance structure of the deployment path, followed by recentering to its mean. The alignment requires only RTT observations from the deployment path and does not depend on workload labels.

The two stages are complementary. Dimensionality reduction removes redundant and noisy components, producing a compact and stable representation. Covariance alignment then adjusts the remaining structure to account for residual distribution differences between paths. Together, they enable consistent feature representations across network environments.

\vspace{-0.085in}

\subsection{Incremental Workload Inference}

A single observation window provides only a limited and noisy view of the underlying workload. Since the workload remains constant over the duration of a run, each window can be interpreted as an independent, noisy observation of the same latent variable. The inference problem therefore reduces to combining evidence from multiple such observations over time.

For each window $j$, a classifier operates on the corresponding feature vector $\mathbf{x}_j$ and produces a posterior probability distribution over the workload classes:
\[
\mathbf{p}_j = [P(y = w_1 \mid \mathbf{x}_j), \ldots, P(y = w_C \mid \mathbf{x}_j)]
\]
where $\mathcal{W} = \{w_1, \ldots, w_C\}$ is the set of candidate workloads. In our system, these posterior estimates are produced by classifiers in the inference pipeline.
To aggregate information across windows, we adopt a probabilistic formulation. Under the assumption that window-level observations are conditionally independent given the workload, the joint likelihood over a sequence of $T$ windows is:
\[
P(\mathbf{x}_{1:T} \mid y = w_c) = \prod_{j=1}^{T} P(\mathbf{x}_j \mid y = w_c)
\]

In practice, classifiers provide posterior estimates $P(y = w_c \mid \mathbf{x}_j)$. Under a uniform class prior, the posterior is proportional to the likelihood, and combining posteriors multiplicatively is equivalent to maximum-likelihood estimation over the full observation sequence. For numerical stability, we perform accumulation in log space, yielding:
\[
L_c = \sum_{j=1}^{T} \log P(y = w_c \mid \mathbf{x}_j)
\]
where $L_c$ represents the accumulated evidence supporting class $w_c$.
The final prediction is:
\[
\hat{y} = \arg\max_c \; L_c
\]

This formulation has several useful properties. It is incremental, allowing the prediction to be updated as new observations arrive. It naturally aggregates weak evidence across windows, enabling reliable inference even when individual observations are noisy. It also provides a measure of confidence through the separation between accumulated scores across classes.

\begin{figure}[tb]
\centering
\includegraphics[width=\columnwidth]{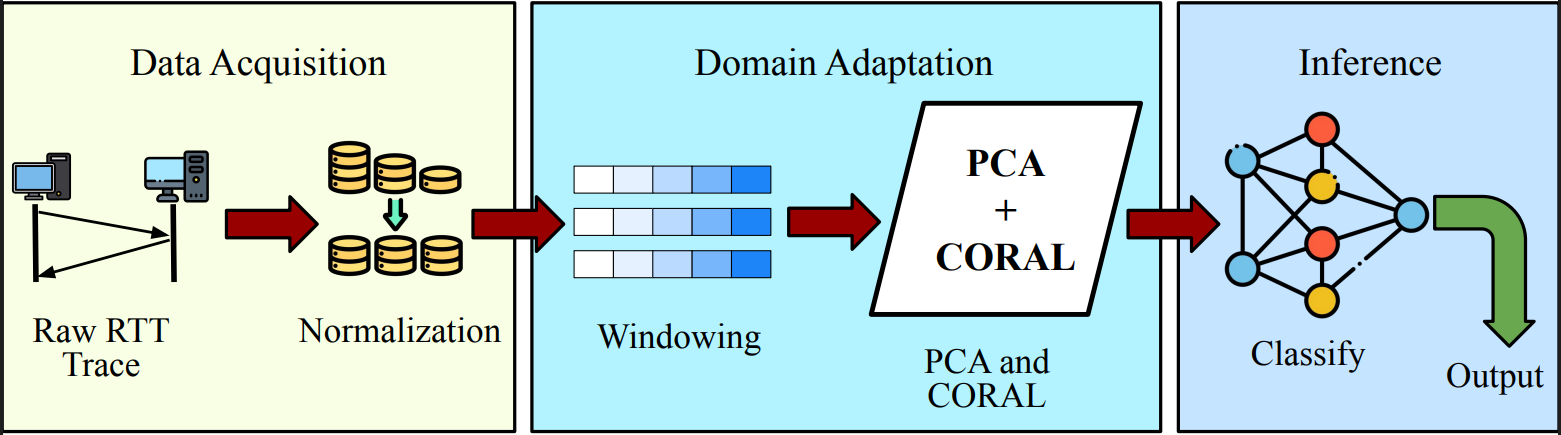}
\caption{Flow diagram of the entire pipeline used for workload inference} 

\vspace{-0.25in}
\label{fig:inference}
\end{figure}

Importantly, a maximally uncertain prediction contributes equally to all classes and therefore does not bias the result, while a confident but incorrect prediction contributes strongly in the wrong direction. Robustness therefore arises from aggregation over multiple windows: as long as correct predictions occur more frequently than incorrect ones, the correct class accumulates evidence faster over time.

This aggregation scheme applies to classifiers that produce calibrated posterior estimates, such as tree-based models and neural networks. Models that do not produce probabilities (e.g., support vector machines) can instead use alternative aggregation schemes such as majority voting, while sequence-based models may directly produce run-level predictions. The overall inference pipeline is illustrated in Figure \ref{fig:inference}.


\vspace{-0.1in}
\section{Evaluation and Results}
\label{sec:eval}

\subsection{Implementation and Experimental Setup}

\begin{figure}[tb]
  \centering
  \includegraphics[width=0.85\columnwidth]{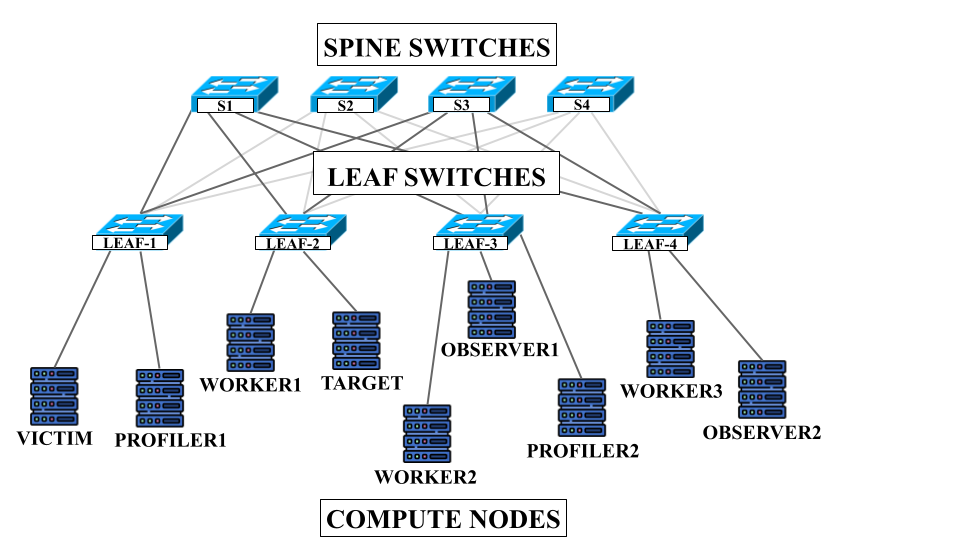}
  \caption{Actual Topology implemented on FARBIC testbed}
  \vspace{-0.2in}
  \label{fig:deployment}
\end{figure}

As shown in Figure \ref{fig:deployment}, we implement and evaluate the framework on the FABRIC testbed \cite{fabric-2019}, using a geographically distributed 4-spine, 4-leaf Clos topology realized with Open vSwitch. The 17-node deployment consists of four spine and four leaf nodes with 4 CPU cores and 8 GB RAM each, three worker and two profiling nodes with 8 cores and 16 GB RAM, two observer nodes with 4 cores and 8 GB RAM, and victim and target nodes with 16 cores and 32 GB RAM, all nodes having 100 GB storage. Host-facing links are rate limited to induce transient queue buildup at leaf-to-spine uplinks. In production datacenters, ECMP may distribute victim traffic across multiple spines, reducing the fraction of traffic visible to any single probe path. We therefore use STP to enforce one active inter-leaf path, creating a controlled best-case setting in which workload and probe traffic share the same spine. The inference principle may also apply under ECMP when their paths overlap, but such settings are outside the scope of this study.

Within this deployment, the \emph{victim} node receives workload traffic generated by \emph{worker} nodes distributed across other leaves, while \emph{profiling} nodes generate labeled workload traffic toward a separate \emph{target} node along a disjoint path. The \emph{observer} nodes are placed on different leaves and probe the victim or target using UDP echo requests without participating in the workload. Under STP, workload and probe traffic traverse the same active spine and bottleneck link, allowing worker-induced congestion to affect the observer's RTT measurements without requiring access to the victim's traffic.

The observer collects RTT samples using lightweight synchronous UDP echo probes, avoiding transport-layer congestion control and retransmission effects so that the measured delay primarily reflects network conditions. We evaluate 15 representative workloads across three categories: distributed communication (allreduce, pipeline, shuffle, federated learning, and scatter-gather), client-server (checkpoint, data preprocessing, database queries, inference serving, log aggregation, microservice communication, and distributed filesystem reads), and bandwidth-intensive workloads (bulk transfer, storage replication, and migration). Multiple workers execute each workload so that traffic converges at the shared spine. Labeled RTT traces from \emph{profiling} paths are used for training, while evaluation uses only RTT observations collected on disjoint \emph{deployment} paths.

The RTT traces are converted into windowed features and processed using the domain-adaptation pipeline in Section \ref{sec:solution-strategy}. We compare models that classify each window independently, including Random Forest, LightGBM, SVM, and MLP, with sequence models, including bidirectional LSTM and Window Transformer, that capture dependencies across windows. All models are trained offline on profiling-path data and evaluated on unseen deployment paths using run-level accuracy, macro-averaged F1-score, and recall.

\vspace{-0.1in}

\subsection{Results}
We present several results in this work based on the data that the observer profiles and the victim path the observer probes.
Models are trained on RTT traces collected from multiple profiling paths and evaluated on a disjoint victim path not seen during training, ensuring that results reflect true cross-path generalization.

\subsubsection{Cross-Path Classification with all 15 workloads: }

\label{sec:result1}

\begin{figure}[tb]
  \centering

  \begin{subfigure}[t]{0.48\columnwidth}
    \centering
    \includegraphics[width=\linewidth]{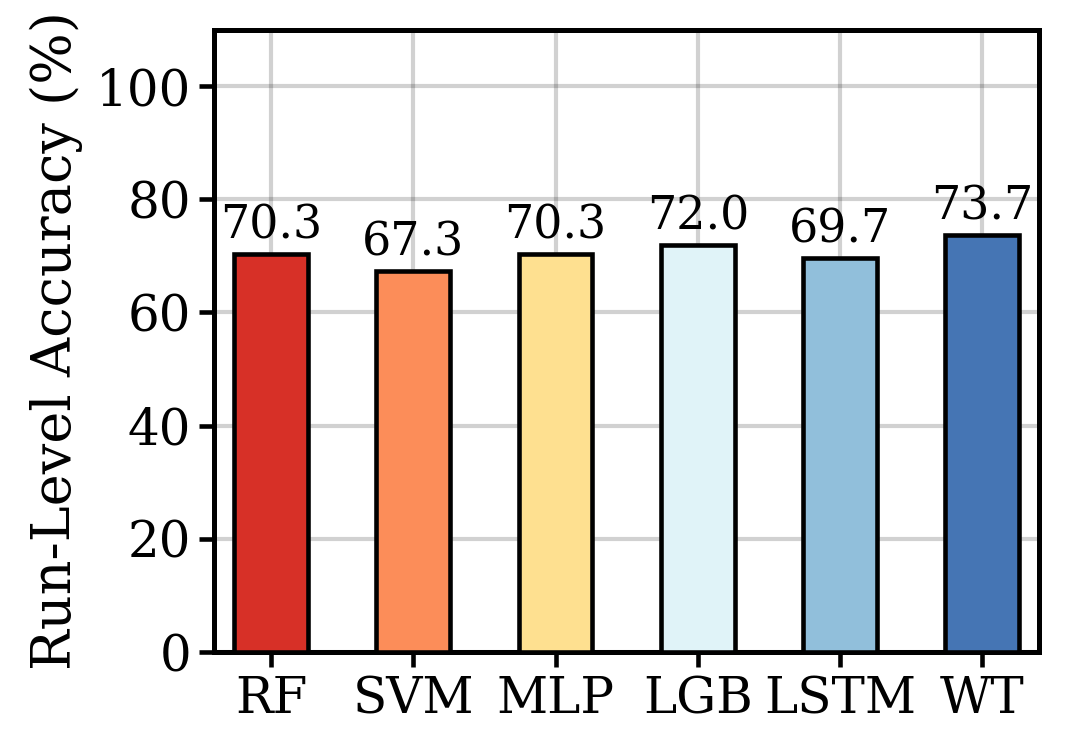}
    \caption{All 15 workloads}
    \label{fig:resultvb-bar}
  \end{subfigure}
  \hfill
  \begin{subfigure}[t]{0.48\columnwidth}
    \centering
    \includegraphics[width=\linewidth]{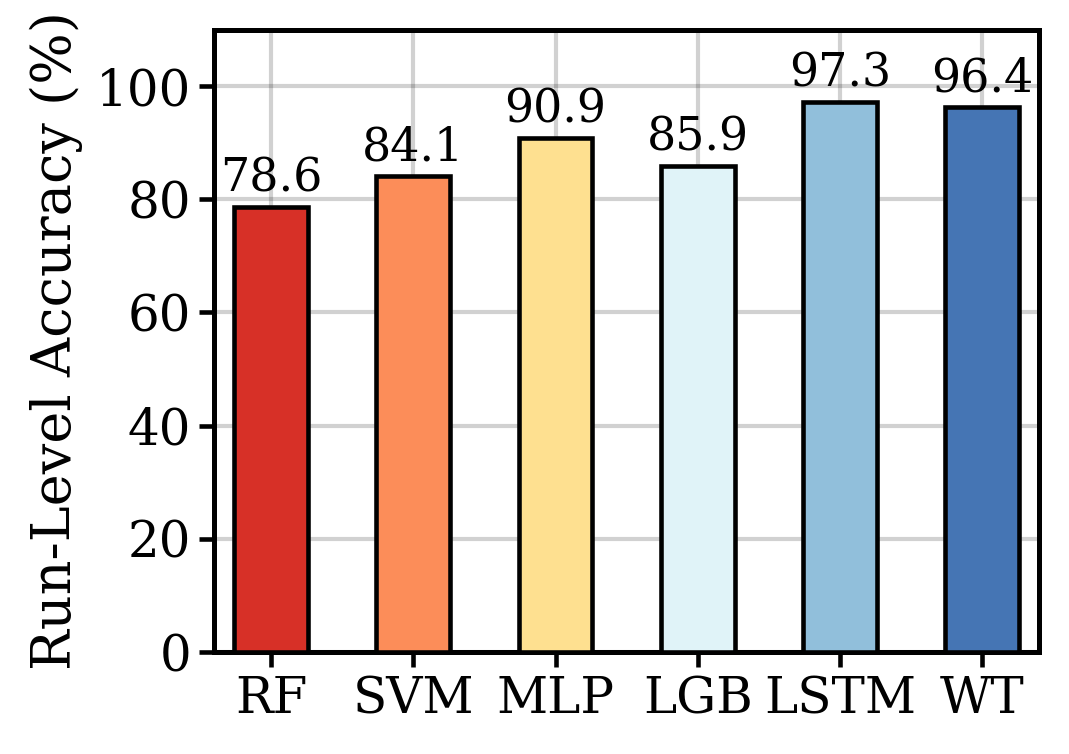}
    \caption{11 workloads}
    \label{fig:resultvd-bar}
  \end{subfigure}

  \caption{Run-level accuracy under different classification models.}
  \label{fig:runlevel-comparison}
  \vspace{-0.1in}
\end{figure}

Figure \ref{fig:resultvb-bar} shows the run-level accuracy across all evaluated models, with the best-performing model achieving 73.7\% accuracy and most models operating in the 67-73\% range. This demonstrates that RTT measurements alone contain sufficient signal to enable meaningful workload inference, even under cross-path variation.

\begin{figure}[tb]
  \centering
  \includegraphics[width=0.75\columnwidth]{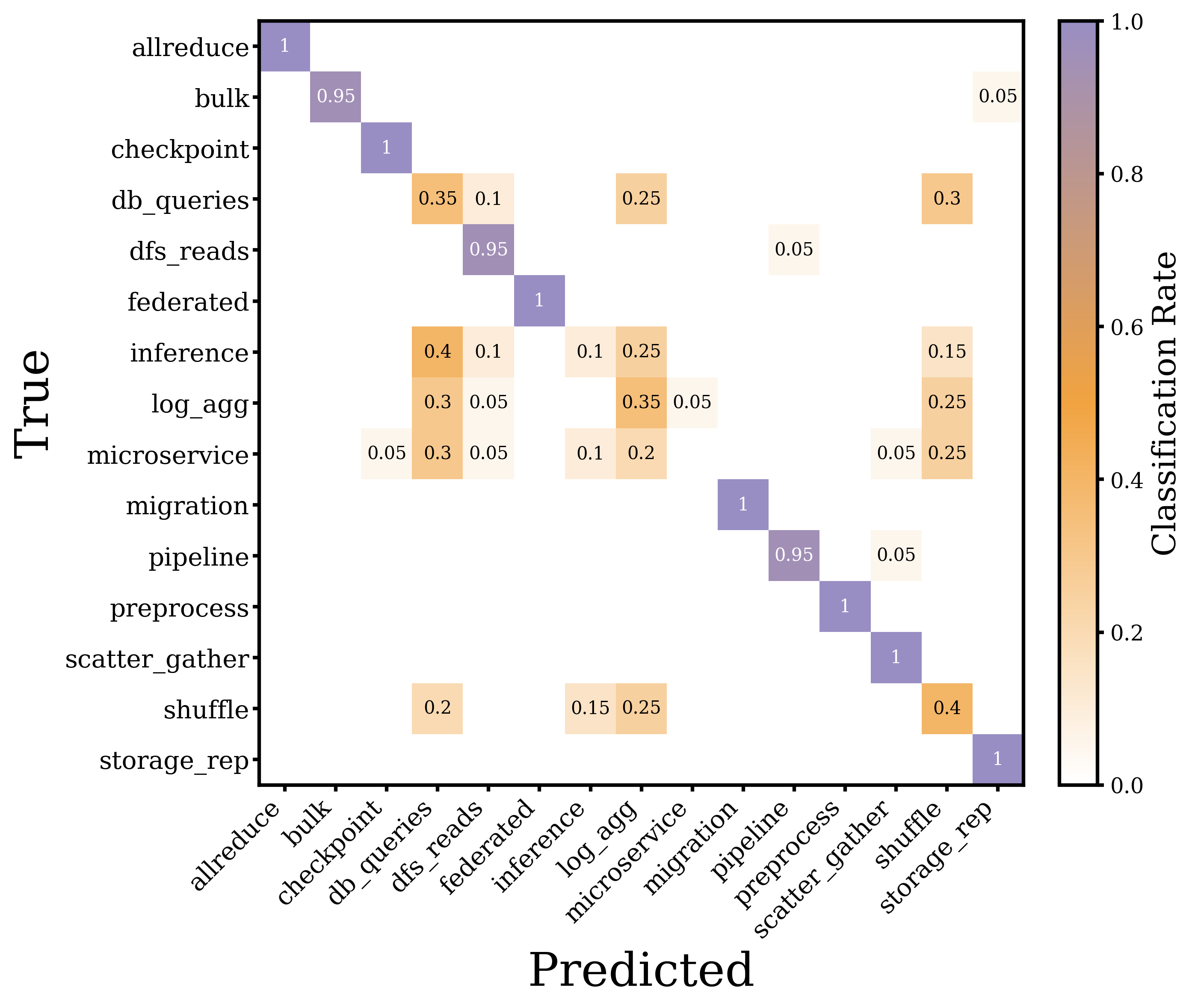}
  \vspace{-0.1in}
  \caption{Confusion Matrix for the classification of 15 workloads through the best performing model - Window Transformer}
  \label{fig:resultvb-confusion}
  \vspace{-0.2in}
\end{figure}

The confusion matrix in Figure \ref{fig:resultvb-confusion} reveals a clear separation in performance across workload types. Ten workloads, including distributed communication and bandwidth-intensive applications, are classified with near-perfect recall (95-100\%), driven by strong and structured congestion signatures. In contrast, five workloads (inference, database queries, microservice traffic, log aggregation, and shuffle) exhibit significantly lower recall (0--40\%), as they generate traffic below the observable noise floor. Misclassifications are almost entirely confined within this low-traffic group, with negligible confusion between high- and low-traffic workloads. This behavior indicates a fundamental detection limit of the RTT side channel rather than a limitation of the classifier.

\subsubsection{Light Traffic  RTT Signal Observability Analysis}

\begin{figure}[tb]
    \centering
    \begin{subfigure}[t]{0.49\linewidth}
        \centering
        \includegraphics[width=\linewidth]{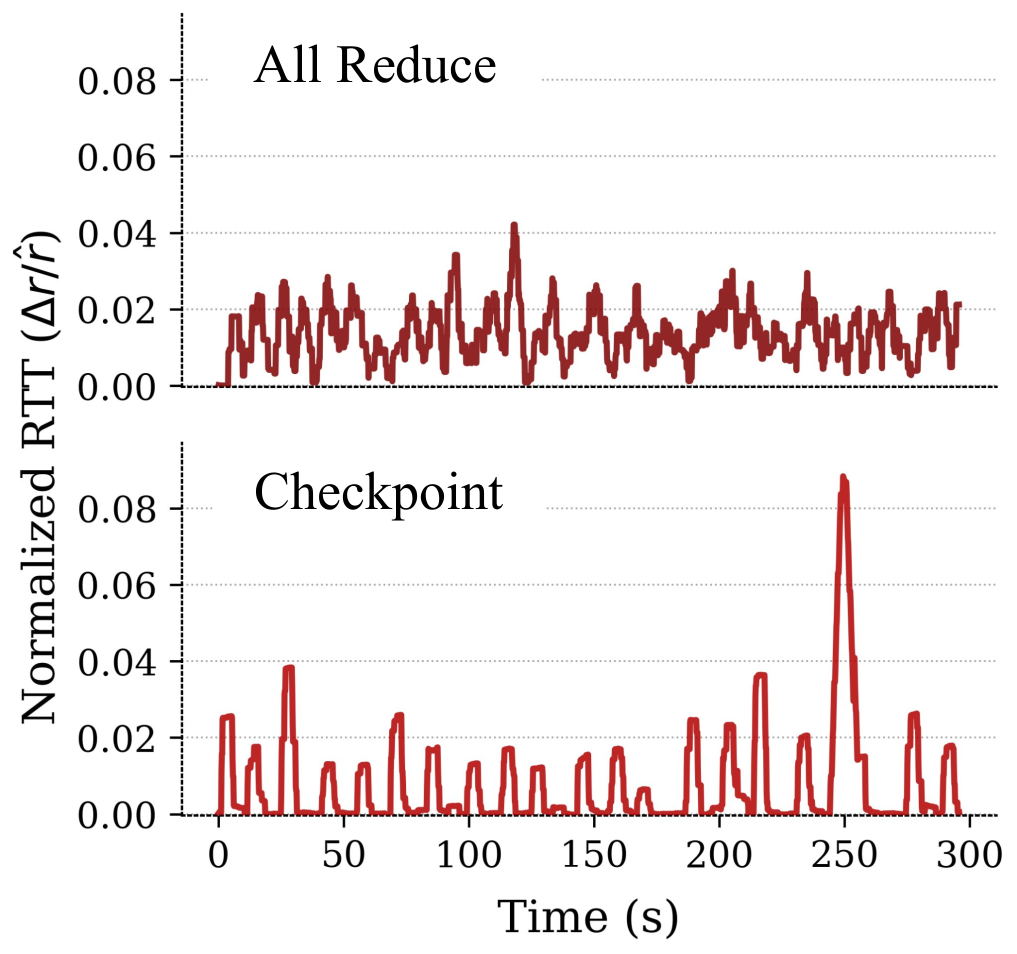}
        \caption{Heavy Traffic}
        \label{fig:result1vca}
    \end{subfigure}
    \hfill
    \begin{subfigure}[t]{0.49\linewidth}
        \centering
        \includegraphics[width=\linewidth]{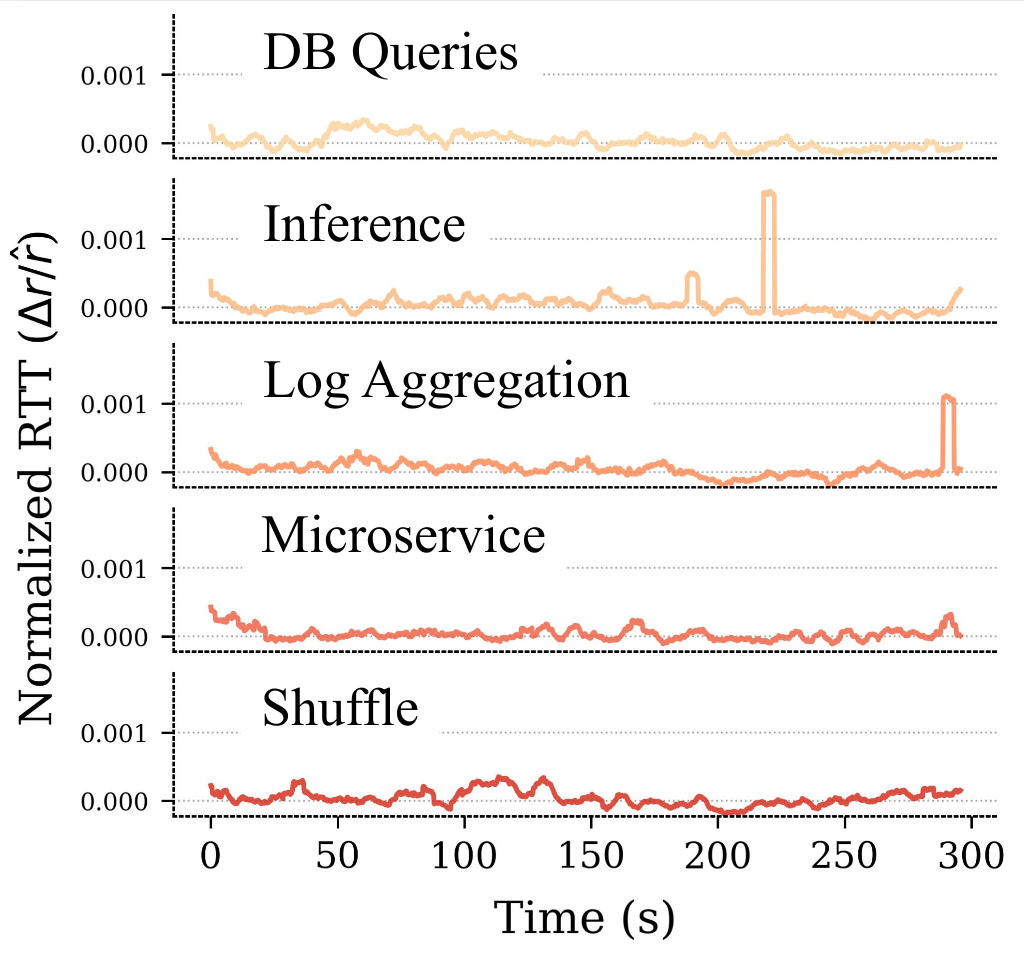}
        \caption{Light Traffic}
        \label{fig:result1vcb}
    \end{subfigure}
    \caption{RTT traces of heavy vs light traffic workloads with 5-second rolling average to suppress probe-level jitter and reveal underlying traffic patterns.
    }
    \label{fig:resultvc}
    \vspace{-0.3in}
\end{figure}

The poor performance observed for five workloads in Section \ref{sec:result1} arises from the absence of a distinguishable congestion signal in their RTT traces. Figure \ref{fig:resultvc} illustrates representative normalized RTT traces for both heavy and light workloads. Heavy workloads such as \emph{allreduce} and \emph{checkpoint} exhibit clear and structured congestion patterns, including periodic bursts and sustained deviations from the baseline. These patterns differ not only in magnitude but also in temporal structure, making them visually and statistically distinguishable. In contrast, light workloads such as database queries, inference, log aggregation, microservice traffic, and shuffle produce RTT traces that remain close to the baseline, with only minor fluctuations and no discernible temporal structure.

Quantitatively, the magnitude of RTT perturbations for light workloads is orders of magnitude smaller than that of heavy workloads, with normalized variations typically around $10^{-3}$ compared to $10^{-2}$ to $10^{-1}$ for heavy traffic. At this scale, the signal falls below the measurement noise floor and becomes indistinguishable across different light workloads. This behavior reflects a fundamental limitation of the RTT side channel; \emph{workloads that do not induce measurable queue buildup cannot be differentiated}. This observation explains the confusion observed in Section \ref{sec:result1} and motivates a reduced-class formulation, where these workloads are grouped into a single \emph{light traffic} class representing the practical observability boundary of the RTT side channel.

\subsubsection{Classification Performance after Class Reduction (11-Class)}

We next evaluate the reduced 11-class formulation under the same cross-path setting as in Section \ref{sec:result1}.
Figure \ref{fig:resultvd-bar} shows the run-level accuracy after merging the five indistinguishable light-traffic workloads into a single class. All models exhibit a substantial and consistent improvement, with performance increasing from the 67-73\% range to 78.6-97.3\%, and the best model achieving 97.3\% accuracy. This improvement is observed across all model families, with sequence models (LSTM and Window Transformer) achieving near-perfect performance. The gains directly reflect the removal of ambiguity among workloads that do not produce a measurable congestion signal.

\begin{figure}[tb]
  \centering
  \includegraphics[width=0.8\columnwidth]{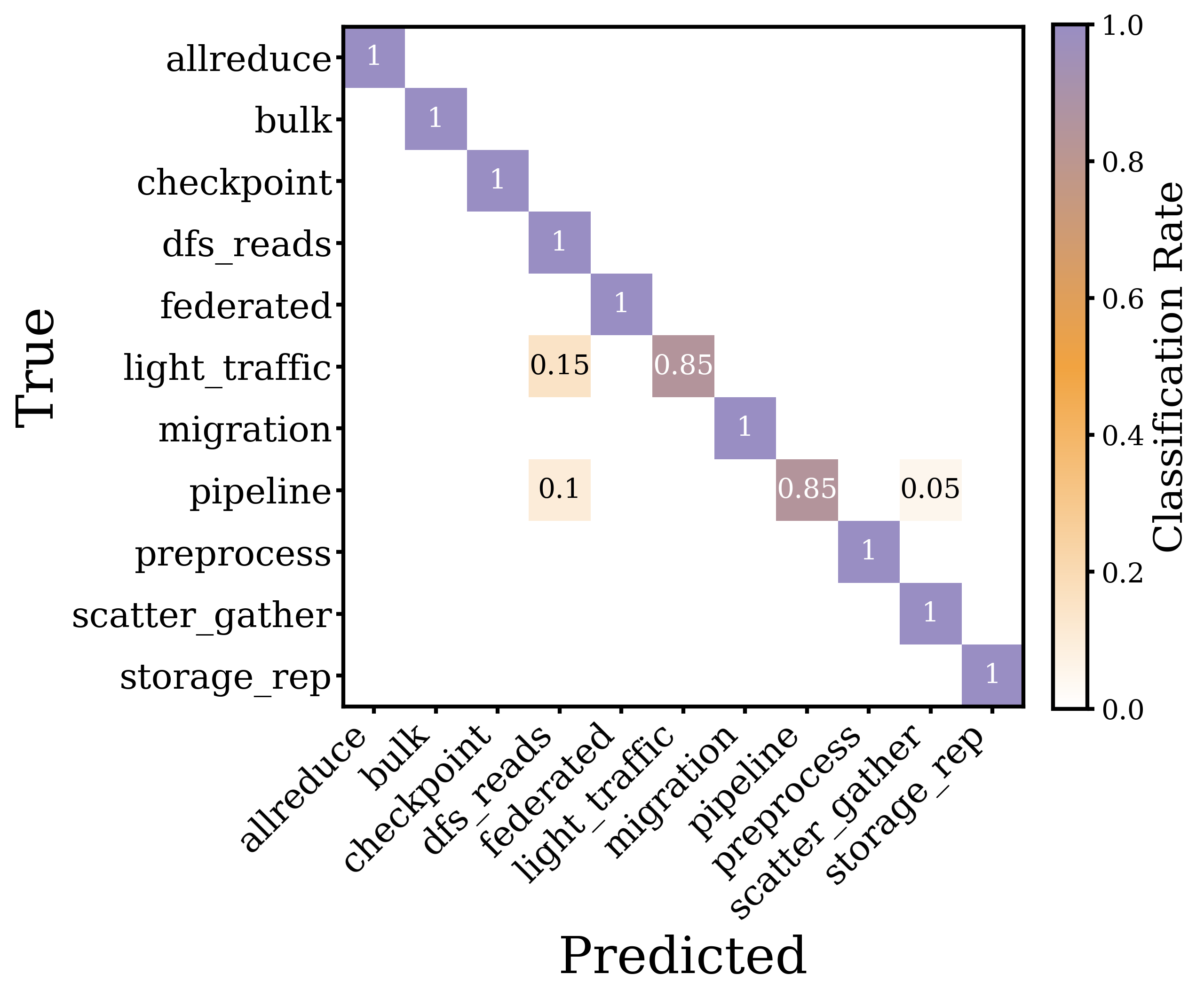}
  \caption{Confusion Matrix for the classification of 11 workloads through the best performing model - Window Transformer}
  \label{fig:resultvd-confusion}
  \vspace{-0.1in}
\end{figure}

The confusion matrix in Figure \ref{fig:resultvd-confusion} is nearly diagonal, confirming that the remaining workload classes are reliably separable through the RTT side channel. Heavy workloads achieve near-perfect precision and recall, while the merged \emph{light\_traffic} class captures residual low-signal behavior with high accuracy. The remaining errors are minor and confined to occasional overlap between low-intensity activity patterns.

\subsubsection{Cross-Path Robustness}

\begin{figure}[tb]
  \centering
  \includegraphics[width=0.9\columnwidth]{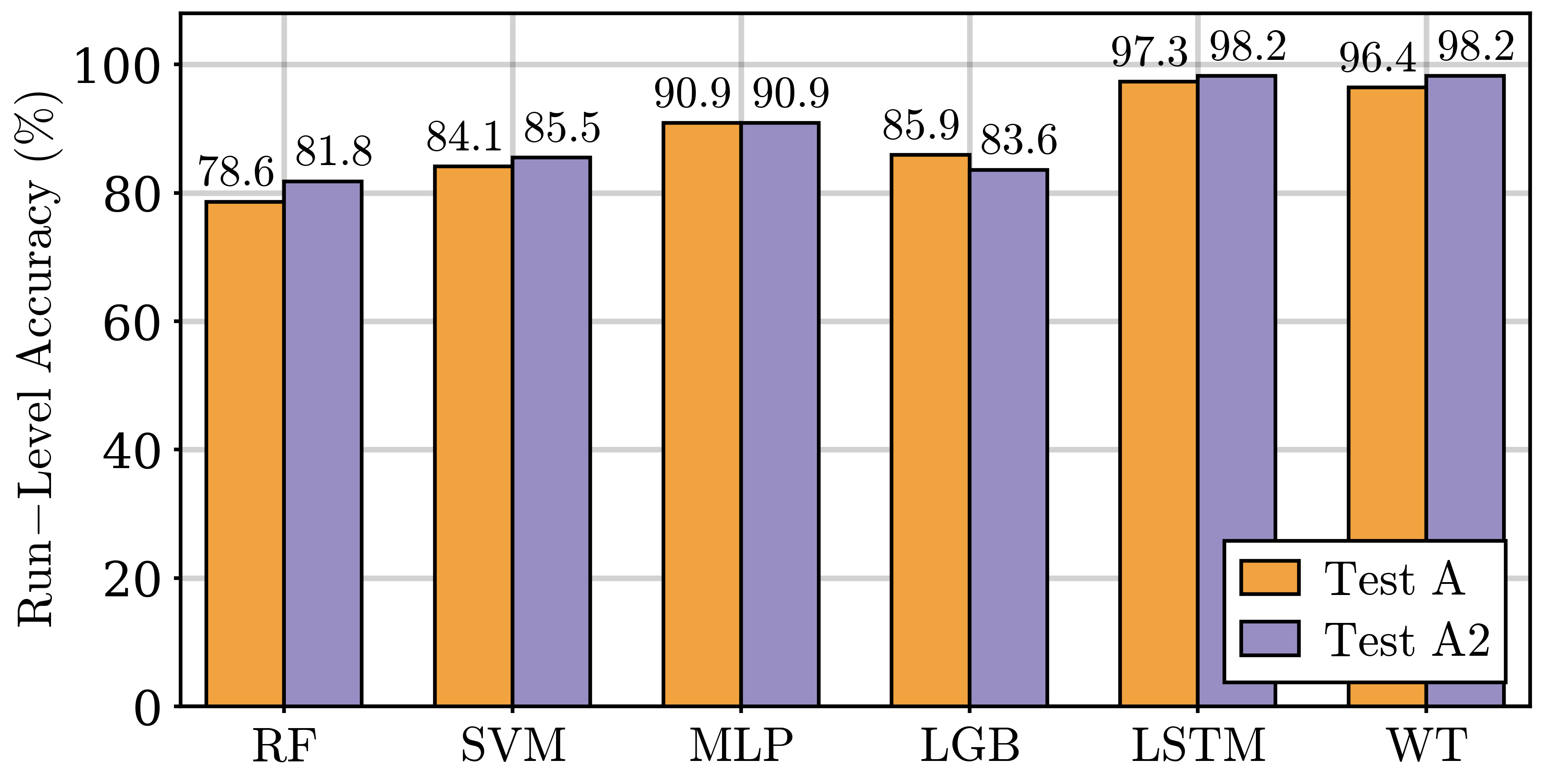}
  \caption{Performance comparison between two different path configurations for victim workloads.}
  \label{fig:resultve}
  \vspace{-0.2in}
\end{figure}

Figure \ref{fig:resultve} shows run-level accuracy across two independently collected victim paths, denoted A and A2. Both correspond to the same observer--victim pair but are collected under different conditions, with A2 consisting of longer runs and a separately collected profiling counterpart. Performance remains consistent across both paths, with all models achieving comparable accuracy and the best models reaching 97-98\%. This consistency indicates that the inference performance is not tied to a specific path instance and generalizes across independently collected victim paths.

Performance on A2 is slightly higher for most models, which can be attributed to longer observation durations that provide more samples for temporal aggregation and improve prediction stability. Across both paths, sequence models (LSTM and Window Transformer) consistently outperform per-window classifiers, achieving near-perfect accuracy. This reinforces that temporal structure in RTT signals is critical for accurate workload inference and remains robust under varying collection conditions.

\subsubsection{Parameter Generalization}

\begin{figure}[tb]
  \centering
  \includegraphics[width=0.7\columnwidth]{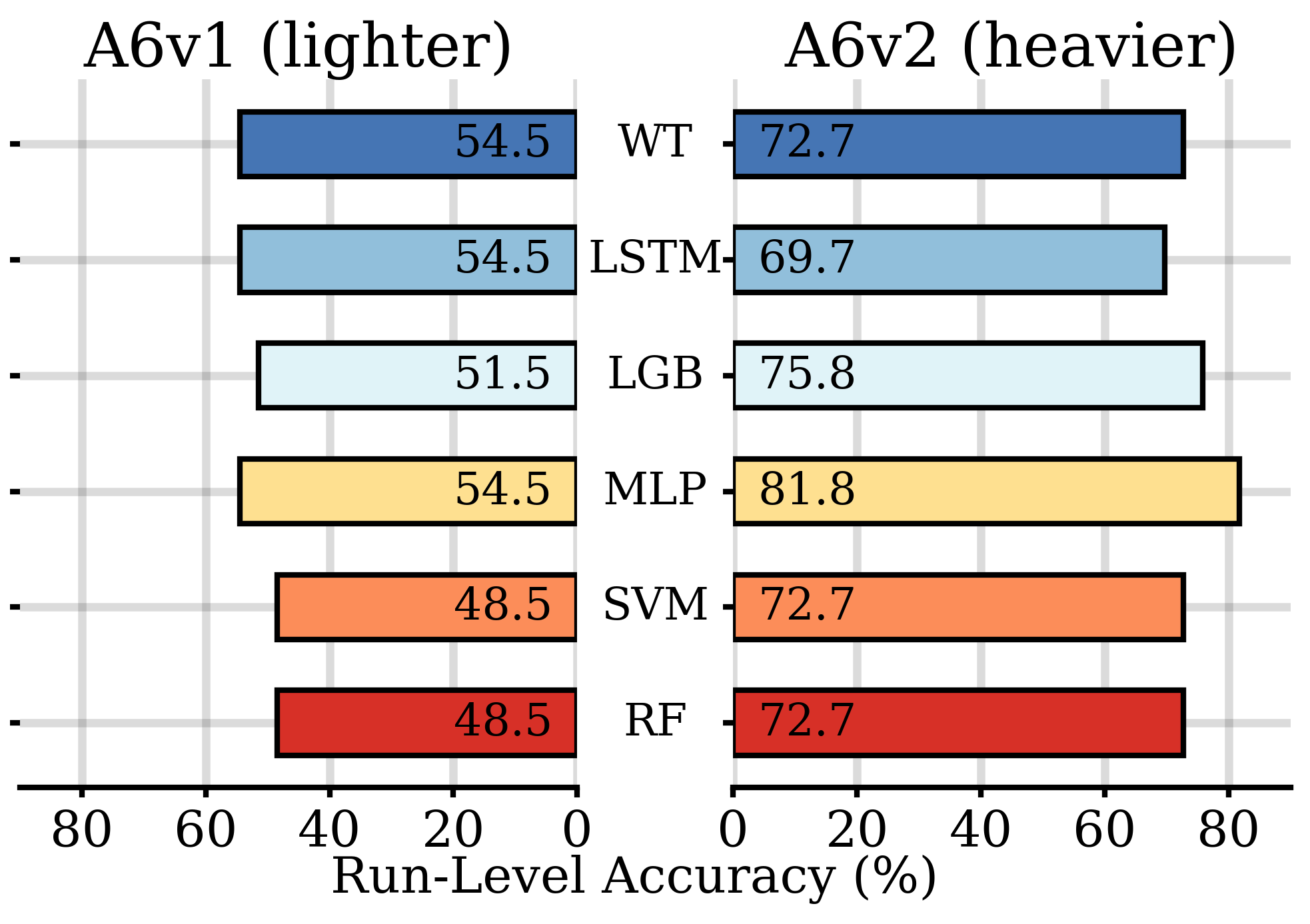}
  \caption{Classification Performance comparison between lighter vs. heavier workloads on the victim}
  \label{fig:resultvf}
  \vspace{-0.2in}
\end{figure}

Figure \ref{fig:resultvf} shows run-level accuracy when workloads are executed with different parameter configurations at test time. Two settings are considered: \emph{A6v1} (lighter workloads with reduced input sizes and traffic intensity) and \emph{A6v2} (comparable or heavier workloads). A consistent performance gap is observed across all models, with accuracy in A6v2 reaching 69-82\% while A6v1 remains limited to 48-55\%.

This behavior arises from the strength of the congestion signal induced by the workloads. When workloads are heavier, they generate stronger traffic bursts that lead to measurable queue buildup at the shared bottleneck, producing clear RTT signatures that match those seen during training. In contrast, lighter workloads generate significantly less traffic and therefore do not induce sufficient congestion to leave a distinct footprint in RTT measurements. As a result, their signatures weaken and become difficult to distinguish from other low-intensity workloads. Across both settings, MLP achieves the highest accuracy, while sequence models exhibit smaller performance gaps between A6v1 and A6v2, indicating slightly better robustness to changes in workload intensity.

\subsubsection{Inference under Mixed Workloads}

\begin{figure}[tb]
  \centering
  \includegraphics[width=0.75\columnwidth]{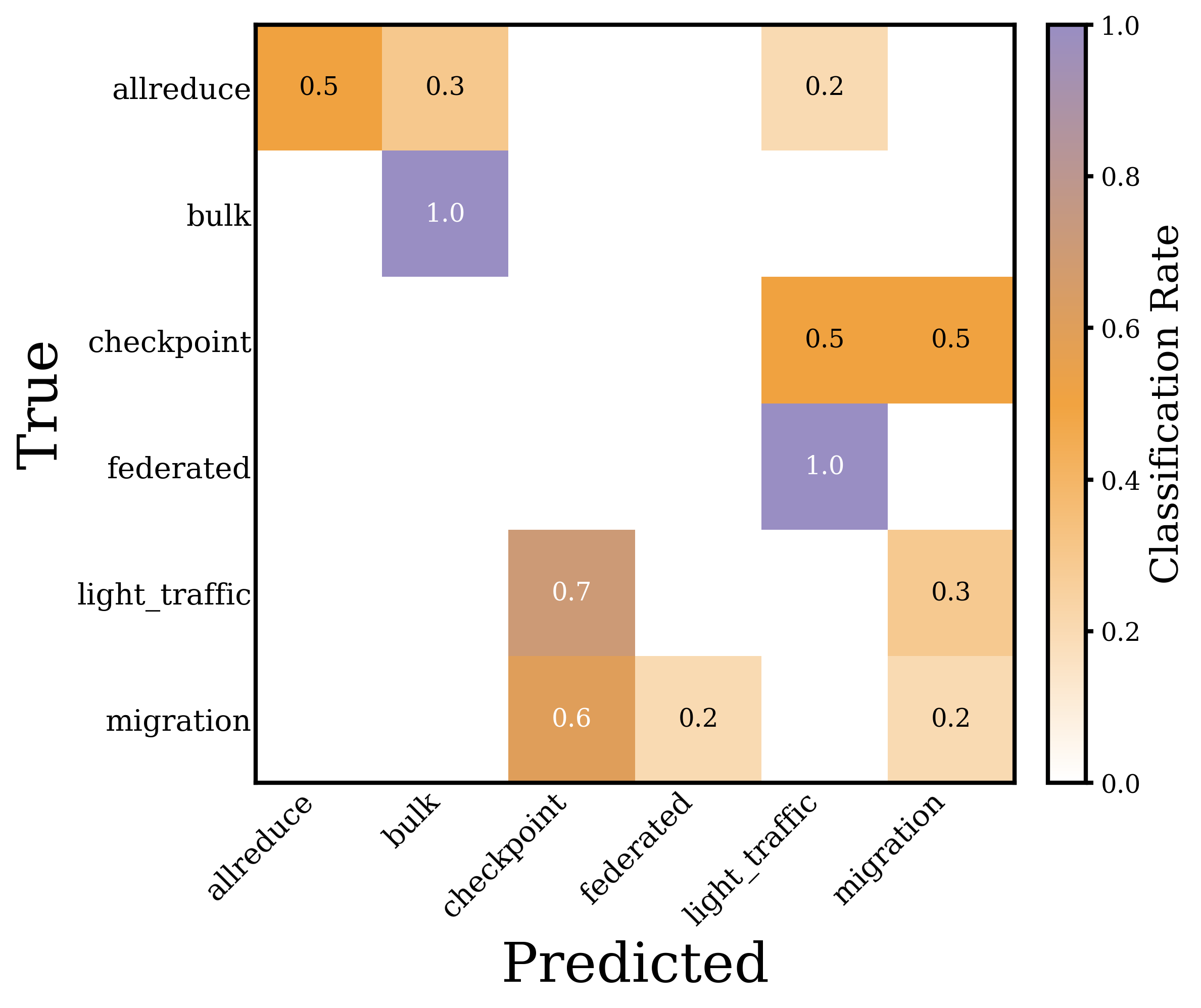}
  \caption{Confusion Matrix for classification of mixed workload scenario}
  \label{fig:resultvg}
  \vspace{-0.2in}
\end{figure}

Figure \ref{fig:resultvg} shows the confusion matrix when multiple workloads are active simultaneously, resulting in overlapping congestion patterns in the RTT signal. Unlike prior experiments where each run corresponds to a single workload, the observed RTT reflects a superposition of traffic from multiple sources, making inference more challenging.

The results indicate that the classifier tends to identify the workload producing the dominant observable congestion signature in the mixture, rather than identifying every concurrently active workload. Workloads with stronger traffic footprints, such as bulk, are classified with high confidence, while others exhibit partial confusion with similar high-intensity patterns. For example, allreduce is sometimes misclassified as bulk or light\_traffic, and checkpoint shows overlap with migration due to similar burst-driven congestion behavior. The \emph{light\_traffic} class remains partially identifiable but shows reduced accuracy due to interference from stronger workloads.

Overall, while accuracy degrades compared to the single-workload setting, the classifier continues to extract meaningful information from the RTT signal. These results suggest that workload information can remain observable under concurrent activity, but the framework’s resolution is limited when multiple workloads generate overlapping congestion signatures.

\subsubsection{Streaming Convergence}

\begin{figure}[tb]
    \centering
    \begin{subfigure}[t]{0.49\linewidth}
        \centering
        \includegraphics[width=\linewidth]{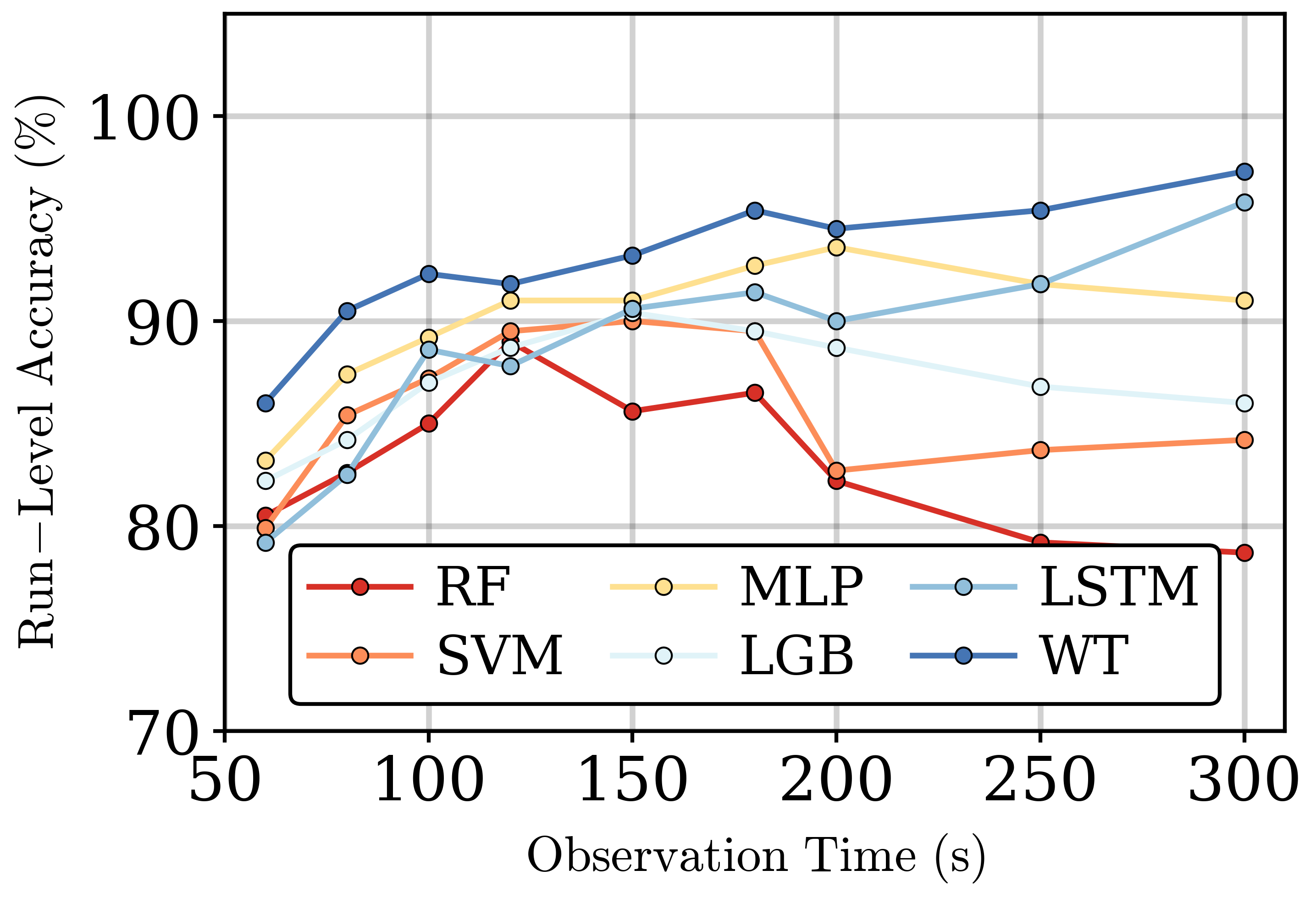}
        \caption{Convergence of all models}
        \label{fig:result1vh-a}
    \end{subfigure}
    \hfill
    \begin{subfigure}[t]{0.49\linewidth}
        \centering
        \includegraphics[width=\linewidth]{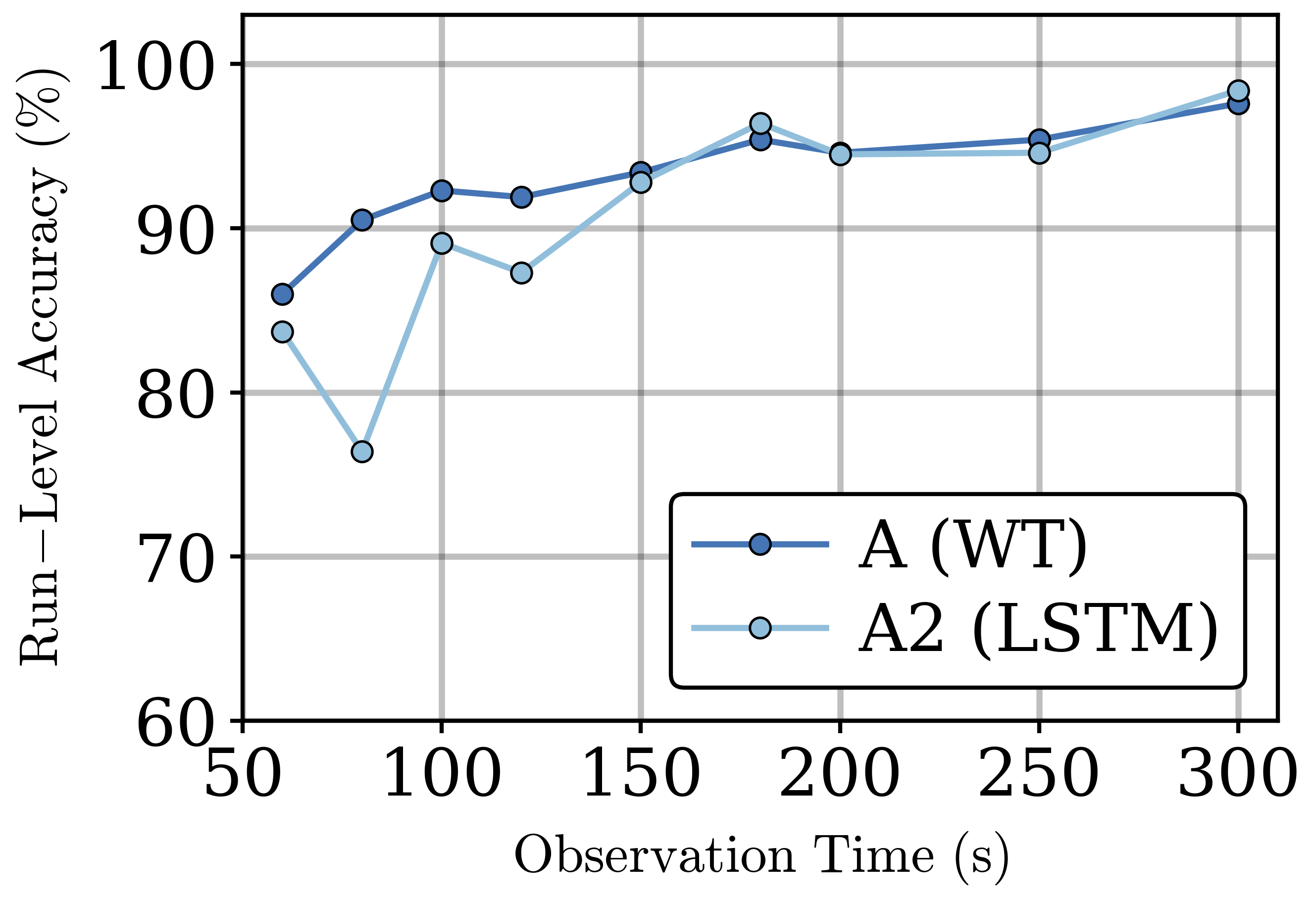}
        \caption{Convergence across different configurations}
        \label{fig:result1vh-b}
    \end{subfigure}
    \caption{Run-level classification accuracy as a function of observation time, illustrating the tradeoff between early inference and convergence to near-optimal performance}
    \label{fig:resultvh}
    \vspace{-0.2in}
\end{figure}

Figure \ref{fig:result1vh-a} shows the convergence behavior of all evaluated models as a function of observation time on test set A, while Figure \ref{fig:result1vh-b} compares the best-performing models across A and A2, where A2 corresponds to a longer-run configuration. Even with a single window (60 seconds), all models achieve 80-86\% accuracy, indicating that the inference framework provides a strong initial estimate with minimal observation. As the observation window increases, accuracy improves for most models, reaching 90\%+ within approximately 100-150 seconds. However, the behavior diverges across model families. Sequence models (LSTM and Window Transformer) steadily improve with longer observations, whereas several per-window classifiers plateau and eventually degrade due to the accumulation of noisier evidence in later windows. MLP remains relatively stable among per-window models, suggesting better probability calibration compared to RF, SVM, and LGB.

Figure \ref{fig:result1vh-b} highlights that the Window Transformer consistently achieves the highest accuracy on A, while LSTM shows comparable performance on A2. Both models exhibit monotonic improvement with increasing observation time, reaching near-peak accuracy (95-98\%) within 180--300 seconds. A2 converges faster than A, reflecting cleaner and more consistent congestion signals under the longer-run configuration.

Overall, these results demonstrate that the inference process is effective in a streaming setting. A useful estimate can be obtained within one minute, while near-optimal accuracy is achieved within a few minutes of observation. The consistent improvement of sequence models further highlights the importance of temporal structure in extracting reliable signals from RTT measurements.

\subsubsection{Realtime-Inference Overhead}

Inference uses 60-second windows with a 10-second stride. As shown in Table \ref{tab:inference_overhead}, the live pipeline incurs approximately 87 ms of end-to-end latency per window, which is small relative to both the observation window and stride and enables predictions shortly after each window closes.

\begin{table}[tb]
\centering
\caption{Per-stage inference latency (ms) measured on the observer node under live deployment}
\label{tab:inference_overhead}

\setlength{\tabcolsep}{4pt}
\renewcommand{\arraystretch}{1.15}

\begin{tabular}{|p{3.35cm}|>{\centering\arraybackslash}p{3.45cm}|}
\hline
\scriptsize\textbf{Stage} & \scriptsize\textbf{Latency (ms)} \\
\hline
\scriptsize Feature extraction & \scriptsize 42.9 \\
\hline
\scriptsize PCA transform & \scriptsize 0.31 \\
\hline
\scriptsize StandardScaler & \scriptsize 0.08 \\
\hline
\scriptsize Random Forest & \scriptsize 30.1 \\
\hline
\scriptsize SVM & \scriptsize 0.9 \\
\hline
\scriptsize MLP & \scriptsize 0.2 \\
\hline
\scriptsize LightGBM & \scriptsize 2.4 \\
\hline
\scriptsize LSTM (3-seed) & \scriptsize 3.0 \\
\hline
\scriptsize WT (5-seed) & \scriptsize 6.9 \\
\hline
\multicolumn{2}{|l|}{\scriptsize\textbf{End-to-end latency}} \\
\hline
\scriptsize Total per window & \scriptsize \textbf{86.8} \\
\hline
\multicolumn{2}{|l|}{\scriptsize\textbf{System context}} \\
\hline
\scriptsize Window size & \scriptsize 60 s \\
\hline
\scriptsize Stride & \scriptsize 10 s \\
\hline
\scriptsize Model load (one-time) & \scriptsize 683 ms \\
\hline
\end{tabular}
\vspace{-0.2in}
\end{table}

Feature extraction is the dominant component, accounting for roughly half of the total latency ($\approx$ 43 ms), due to the computation of statistical, spectral, and temporal features. Among classifiers, Random Forest contributes the highest inference cost ($\approx$ 30 ms), while neural models such as LSTM and the Window Transformer remain efficient ($\approx$ 3–7 ms) due to optimized matrix operations. The domain adaptation steps, including PCA and scaling, add negligible overhead ($\textless$ 1 ms). Overall latency remains consistent across workloads, indicating that inference cost is independent of traffic patterns.
These results show that the computational cost of the inference pipeline is small relative to the observation interval.


All datasets, feature extraction scripts, and a detailed README are provided as a GitHub repository \cite{google-drive}.


\subsection{Scope, Assumptions and Limitations}
Our findings establish the feasibility of inferring workload activity from shared-path RTT observations in a controlled datacenter environment. The analysis assumes prior knowledge of the candidate workload set and sufficient contention for workload-dependent temporal patterns to emerge in the observed signal. Inference performance may vary when traffic intensity differs from the profiled conditions, when multiple workloads produce overlapping signatures, or when the induced delay is too weak to distinguish reliably. The experiments use STP to maintain consistent path overlap, while ECMP and other multipath routing policies are not evaluated. These considerations define the scope within which the reported results should be interpreted, without limiting the broader applicability of the underlying inference principle.


\section{Conclusions and Future Work}
\label{sec:conclusion}

This paper investigated a network side-channel vulnerability that enables inference of datacenter workload activity using only RTT measurements. By probing a shared leaf--spine fabric with lightweight UDP packets, the observer captures congestion-induced temporal patterns without access to packet contents or flow metadata. Under cross-path evaluation, the framework achieves 73.7\% run-level accuracy across all workload classes, increasing to 97.3\% when workloads with weak congestion signatures are grouped into a single low-signal class. Predictions improve with longer observation, while the full inference pipeline completes in 87 ms per window, supporting real-time operation with low computational overhead.

Our analysis also identifies the conditions that shape inference performance. Low-intensity workloads may remain below the RTT noise floor, changes in workload intensity can reduce generalization, and concurrent workloads can bias predictions toward the dominant observable signature. These findings suggest that reducing persistent congestion coupling, limiting the observability of delay variation, or disrupting the temporal stability of workload-induced patterns could mitigate such leakage. Future work will examine multipath routing environments, broader workload variation, richer topologies, and the effectiveness and cost of these mitigation strategies.


\bibliographystyle{ieeetr}
\bibliography{refs}

\end{document}